\documentclass{aa}
\usepackage{graphicx}
\usepackage{amssymb}
\begin{document}

\authorrunning{K\"apyl\"a et al.}
\titlerunning{Strouhal number for convection}

   \title{Local models of stellar convection III:}

   \subtitle{The Strouhal number}

   \author{P. J. K\"apyl\"a
	  \inst{1}$^{,}$\inst{2},
          M. J. Korpi
	  \inst{3},
	  M. Ossendrijver
	  \inst{2},
          \and
          I. Tuominen
	  \inst{1}$^{,}$\inst{3},
	  }

   \offprints{P. J. K\"apyl\"a\\
	  \email{petri.kapyla@oulu.fi}
	  }

   \institute{Astronomy Division, Department of Physical Sciences,
              PO BOX 3000, FIN-90014 University of Oulu, Finland
	  \and Kiepenheuer--Institut f\"ur Sonnenphysik, 
	      Sch\"oneckstrasse 6, D--79104 Freiburg, Germany
          \and Observatory, PO BOX 14, FIN-00014 University of Helsinki, 
              Finland \\ }

   \date{Received 25 October 2004 / Accepted 15 November 2005}

   \abstract
     {}
     {The Strouhal number (St), which is a nondimensional measure of 
     the correlation time, is determined from numerical models of 
     convection. The Strouhal number arises in the
     mean-field theories of angular momentum transport and magnetic
     field generation, where its value determines the validity of
     certain widely used approximations, such as the first order
     smoothing (hereafter FOSA). More specifically, the relevant
     transport coefficients can be calculated by means of a cumulative
     series expansion if St $<$ St$_{\rm crit} \approx 1$.}
     {We define the Strouhal number as the ratio of the correlation 
     and turnover times, which we determine separately, the former 
     from the autocorrelation of velocity, and the latter by following 
     test particles embedded in the flow.}
     {We find that the Strouhal numbers are, generally, of the order 
     of 0.1 to 0.4 which is close to the critical value above which 
     deviations from FOSA become significant. Increasing the rotational 
     influence tends to shorten both timescales in such a manner that 
     St decreases. However, we do not find a clear trend as function 
     of the Rayleigh number for the parameter range explored in the 
     present study.}
     {}


   \maketitle

\section{Introduction}
  The mean-field theories of angular momentum transport
  (e.g. R\"udiger \cite{Rudiger80}, \cite{Rudiger89}) and
  hydromagnetic dynamos (e.g. Steenbeck \& Krause \cite{SteenKrau69};
  Krause \& R\"adler \cite{KraRad80}) require knowledge of the
  Reynolds stresses and the mean electromotive force,
  respectively. Direct analytical calculation of these quantities for
  astrophysical purposes is not possible at present due to the lack of
  an established theory of turbulence. Numerical calculations are not
  much better off since in most astrophysical systems the
  computational resources needed in order to resolve all physically
  relevant scales are several orders of magnitude larger than
  currently available.

  In mean-field theories these problems are circumvented by relating
  the Reynolds stresses and electromotive force to the mean quantities
  (the rotation vector $\vec{\Omega}$ and the mean magnetic field
  $\langle \vec{B} \rangle$, respectively) and their gradients by
  means of a cumulative series expansion (see van Kampen
  \cite{vanKampen1974a},\cite{vanKampen1974b}). In the best known and
  most often used approach, the first order smoothing approximation
  (FOSA), only the first terms of these expansions are taken into
  account. This approach can be shown to be valid if either the
  relevant Reynolds number, ${\rm Re} = ul/\nu\;, {\rm Rm} = ul/\eta$,
  or the Strouhal number
  \begin{equation}
    {\rm St} = \frac{|(\vec{u} \cdot \nabla)\vec{u}|}{|\partial \vec{u}/\partial t|} \approx u \frac{\tau_{\rm c}}{l_{\rm c}} \propto \frac{\tau_{\rm c}}{t_{\rm to}}\;,
    \label{equ:st}
  \end{equation}
  is sufficiently small. Above, $u$ and $l$ are the typical velocity
  and length scales, and $\tau_{\rm c}$ and $l_{\rm c}$ are the
  correlation time and length of the turbulence, respectively. We show
  in Sect.\,\ref{subsec:turnovert} that $1/t_{\rm to} \propto u/l_{\rm
    c}$. Only the case of small St is relevant in astrophysical
  circumstances, where typically Re, Rm $\gg$ 1. Hence the condition
  St $\ll$ 1 should be fulfilled if FOSA is to be valid in stellar
  convection zones. If this is not the case then FOSA is likely to be
  too crude a truncation but, as long as St is smaller than the
  critical value for convergence it is possible to construct
  higher-order mean-field theories by including more terms of the
  cumulant expansion. However, a universal critical value of St for
  convergence of the cumulant expansion does not exist because this
  depends on the geometry of the problem and on the flow itself. In
  general though, the critical value is thought to be less than unity.
  For simple turbulence models it is possible to compute it; Nicklaus
  \& Stix (\cite{NickStix88}, hereafter NS88) obtained $\mbox{St}_{\rm
    crit}=1$ for their model. Furthermore, the results of NS88 show
  that the higher order effects remain small in comparison to the FOSA
  result when St $\lesssim 0.5$.

  It is well-known that the requirement for a small Reynolds number is
  not satisfied in stellar environments. However, the question of the
  Strouhal number is not settled. On account of the observations of
  the solar surface granulation, one can estimate the correlation and
  turnover times to be roughly equal, indicating that St $\approx 1$
  at the solar surface (see e.g. Chapter 6 of Stix
  \cite{Stix02}). Similar values can also be estimated for
  supergranulation for which typical numbers are $u \approx
  100$~m~s$^{-1}$, $l \approx 10^7$m, and $\tau \approx
  10^5$~s. However, even for granulation the precise value has, to our
  knowledge, not been established, and nothing is known about St in
  the deeper layers. Furthermore, recent results from forced
  turbulence calculations indicate that if the higher order
  correlations in the equations of the passive scalar flux
  (Brandenburg et al. \cite{Brandea04}, hereafter BKM) and
  electromotive force (Brandenburg \& Subramanian \cite{BranSubr05})
  are taken into account via the so-called minimal
  $\tau$-approximation (Blackman \& Field \cite{BlackField2002},
  \cite{BlackField2003}), the Strouhal number can be seen to
  substantially exceed unity and roughly equal to unity, respectively.

  Although forced turbulence is rather different in comparison to
  convection, the aforementioned studies still raise the question
  whether the results of the convection calculations can be
  interpreted within the framework of the standard mean-field theory
  as has been done in numerous studies during recent years
  (e.g. Brandenburg et al. \cite{Brandea90}; Pulkkinen et
  al. \cite{Pulkkiea93}; Ossendrijver et al. \cite{Ossea01},
  \cite{Ossea02}; K\"apyl\"a et al. \cite{Kapyea04}; R\"udiger et al.
  \cite{Ruedigerea2005}). Motivated by the unknown status of the
  Strouhal number for convection and the previous studies on the
  subject in different contexts, we set out to calculate St from
  numerical calculations of convection. In order to do this, we
  calculate the correlation time from the autocorrelation of velocity
  and determine the turnover time by following test particles embedded
  into the flow and define the Strouhal number as the ratio of the
  two.

  The remainder of the paper is organised as follows:
  Sect.\,\ref{sec:model} summarises briefly the numerical model used
  and in Sect.\,\ref{sec:results} the results of the study are
  discussed. Finally, Sect.\,\ref{sec:conclu} gives the conclusions.

\section{The model}
\label{sec:model}

  A detailed description of the convection model can be found in
  K\"apyl\"a et al. (\cite{Kapyea04}, hereafter Paper I) and the
  convection calculations are made with a setup identical to that used
  in Paper I. See Table \ref{tab:convruns} for a summary of the main
  parameters.

  For the purposes of the present study we have added the possibility
  to follow the trajectories of Lagrangian test particles in the
  model. In order to integrate the trajectory we need the
  velocity at the position of the test particle at each integration
  step. This is done by finding the grid points $(n_x,n_y,n_z)$ next
  to the test particle and using linear interpolation to obtain the
  velocity at the correct position
  \begin{eqnarray}
    \vec{u}(\vec{x}_{\rm tp}) = \vec{u}(\vec{n}) - \sum_{i=x,y,z} \frac{\delta x_i}{\Delta x_i} \vec{u}(\vec{n}-\vec{e}_i)\;, 
  \end{eqnarray}
  where $\vec{x}_{\rm tp} = (x_{\rm tp},y_{\rm tp},z_{\rm tp})$ is the
  position vector of the test particle, $\vec{n}=(n_x,n_y,n_z)$
  denotes the grid point next to the test particle, $\vec{e}_i$ the
  unit vector in direction $i$, $\delta x_i = x_i -x_{\rm tp}$ the
  distance between the test particle and the grid point next to it,
  and $\Delta x_i$ the grid spacing in direction $i$. In the present
  calculations we follow one thousand test particles that are
  introduced at the middle of the convectively unstable layer at
  random horizontal positions after convection has reached a
  statistically stationary state.

  The calculations were made with a modified version of the numerical
  method described in Caunt \& Korpi (\cite{CauKo01}). The
  calculations were carried out on the KABUL and BAGDAD Beowulf
  clusters at the Kiepenheuer-Institut f\"ur Sonnenphysik, Freiburg,
  Germany, and on the IBM eServer Cluster 1600 supercomputer hosted by
  CSC Scientific Computing Ltd., in Espoo, Finland.

  \begin{table}
    \centering
    \caption[]{Summary of the calculations and the main
      parameters. From left to right: the Rayleigh, Reynolds, Taylor,
      and Coriolis numbers, the latitude, and the grid size.}
    \vspace{-0.75cm}
    \label{tab:convruns}
     $$
    \begin{array}{p{0.17\linewidth}cccccc}
      \hline
      \noalign{\smallskip}
      Run       & {\rm Ra} & {\rm Re} & {\rm Ta} & {\rm Co} & \Theta & {\rm Grid} \\
      \noalign{\smallskip}
      \hline
      lCo0      & 1.25\cdot 10^{5}&  95 & 0                 &    0 & -        & 48^3 \\
      Co0       & 2.5\cdot 10^{5} & 140 & 0                 &    0 & -        & 64^3 \\
      mCo0      & 5.0\cdot 10^{5} & 190 & 0                 &    0 & -        & 96^3 \\
      hCo0      & 10^{6}          & 246 & 0                 &    0 & -        & 128^3 \\
      \hline
      Co01-00   & 2.5\cdot 10^{5} & 140 & 203               & 0.10 & \;\; 0 \degr & 64^3\\
      Co01-30   & 2.5\cdot 10^{5} & 140 & 203               & 0.10 & -30 \degr & 64^3\\
      Co01-60   & 2.5\cdot 10^{5} & 138 & 203               & 0.10 & -60 \degr & 64^3\\
      Co01-90   & 2.5\cdot 10^{5} & 138 & 203               & 0.10 & -90 \degr & 64^3\\
      Co1-00    & 2.5\cdot 10^{5} & 139 & 2.03 \cdot 10^{4} & 1.04 & \;\; 0 \degr & 64^3\\
      Co1-30    & 2.5\cdot 10^{5} & 145 & 2.03 \cdot 10^{4} & 1.00 & -30 \degr & 64^3\\
      Co1-60    & 2.5\cdot 10^{5} & 141 & 2.03 \cdot 10^{4} & 1.03 & -60 \degr & 64^3\\
      Co1-90    & 2.5\cdot 10^{5} & 139 & 2.03 \cdot 10^{4} & 1.05 & -90 \degr & 64^3\\
      Co10-00   & 2.5\cdot 10^{5} & 337 & 2.03 \cdot 10^{6} & 4.24 & \;\;0 \degr & 96^2 \times 64\\
      Co10-30   & 2.5\cdot 10^{5} & 121 & 2.03 \cdot 10^{6} & 11.8 & -30 \degr & 96^2 \times 64\\
      Co10-60   & 2.5\cdot 10^{5} & 105 & 2.03 \cdot 10^{6} & 13.6 & -60 \degr & 96^2 \times 64\\
      Co10-90   & 2.5\cdot 10^{5} & 104 & 2.03 \cdot 10^{6} & 13.7 & -90 \degr & 96^2 \times 64\\
      \noalign{\smallskip}
      \hline
    \end{array}
    $$ 
  \end{table}


\section{Results}
\label{sec:results}

  \begin{figure}
  \centering
  \includegraphics[width=0.5\textwidth]{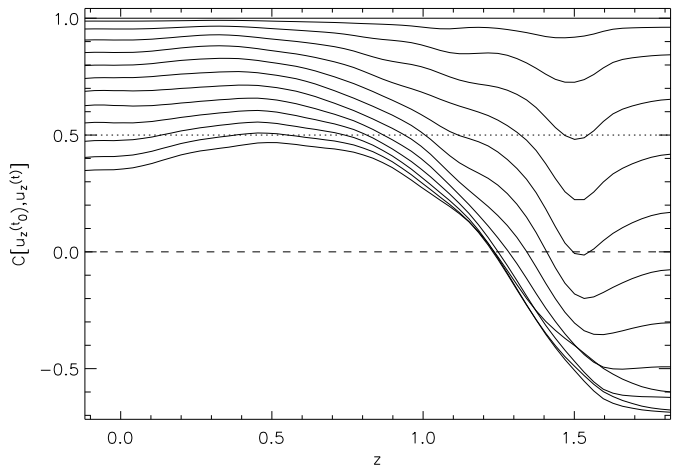}
     \caption{Horizontally averaged correlations of $u_{\rm z}$ with
       respect to the snapshot at ${t = 100}$ for the run Co0. Each
       curve is separated by $\Delta t = 1$ in time units of
       $\sqrt{d/g}$.}
     \label{fig:corr}
  \end{figure}

\subsection{Correlation time}
\label{subsec:corrt}
  We determine the correlation time from the velocity autocorrelation
  function
  \begin{eqnarray}
    C[u_{i}(\vec{x},t_0),u_{i}(\vec{x},t)] = \frac{\langle u_{i}(\vec{x},t_0)
      u_{i}(\vec{x},t)\rangle}{\sqrt{\langle u_{i}^2(\vec{x},t_0) \rangle
        \langle u_{i}^2(\vec{x},t)\rangle}}\;, \label{equ:corr}
  \end{eqnarray}
  where $i$ denotes the velocity component used, $t_0$ and $t$ are the
  times from which the snapshots were taken, and the brackets denote
  horizontal averaging. We estimate the correlation time, $\tau_{\rm
    c}$, to be the time after which the correlation drops below a
  fixed threshold value, in this case 0.5. However, the correlation
  time still depends on depth, and there are discrete time intervals
  between the stored snapshots (see Figure \ref{fig:corr}). To remedy
  the latter, we calculate the correlation time within the convection
  zone for each depth and use linear interpolation to find a more
  accurate value for the time when $C = 0.5$. Furthermore, it makes
  sense to average the correlation time over the convectively unstable
  layer since convection is more of a global rather than local nature
  with the present parameters (see Figure \ref{fig:etestp}). To check
  the time dependence of $\tau_{\rm c}$ we calculate it with respect
  to about two hundred snapshots for each calculation. The final
  correlation time is an average of these individual values.

  Figure \ref{fig:corr} gives an example from run Co0, showing the
  horizontally averaged correlations of the vertical velocity $u_z$
  from twelve snapshots, each separated by one time unit, with respect
  to the snapshot at $t_0 = 100$. The correlation diminishes
  monotonically as a function of time and for the eleventh snapshot
  the correlation is below 0.5 in the whole convection zone. Using the
  procedure described above, we find the correlation time to be
  $\tau_{\rm c} \approx 9.8$ for this snapshot (average over the
  correlation times with respect to 206 different snapshots gives a
  value 9.4, see the second column of Table \ref{tab:strouhal}). If
  the correlation is calculated for one of the horizontal velocity
  components, $\tau_{\rm c}$ is similar in magnitude, except near the
  surface and in a layer immediately below the convection zone where
  it is somewhat longer than the one calculated from the vertical
  velocity. This effect can be understood to arise from the persistent
  horizontal flows near the boundaries of the convection zone where
  the up- and downflows diverge to the horizontal directions. In what
  follows we estimate the correlation time from the vertical velocity,
  for which $\tau_{\rm c}$ remains more or less constant within the
  convection zone as indicated by Figure \ref{fig:corr}. As stated
  above, our final result is an average over snapshots and the
  corresponding standard deviations of the correlation times are given
  in Table \ref{tab:strouhal}.

\subsection{On the correlation length}
\begin{figure}
  \centering
  \includegraphics[width=0.5\textwidth]{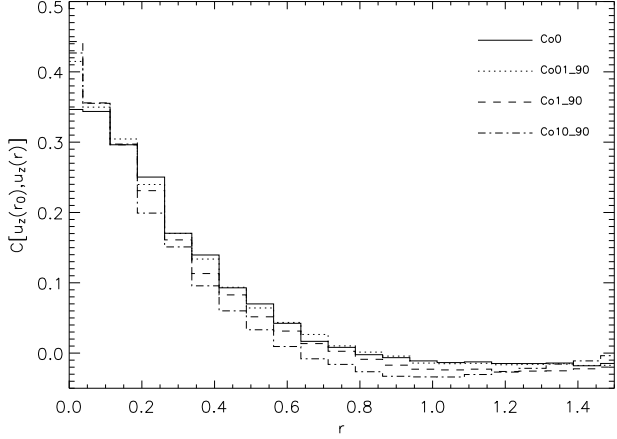}
     \caption{Velocity autocorrelations for $u_{\rm z}$ as functions
       of $r$ and rotation for the runs indicated in the legend.}
     \label{fig:lcorr}
  \end{figure}

  In principle, the correlation length of the turbulence can be
  determined from Eq.~(\ref{equ:corr}) using $r = |\vec{x}_1 -
  \vec{x}_2|$ as the argument instead of the time. In order to do
  this, we choose one thousand random grid points within the
  convectively unstable layer and compute the crosscorrelations of the
  velocities between all the points. We do this procedure for the same
  snapshots from which the correlation times were calculated, and
  average over all snapshots. Due to the finite spatial resolution of
  the calculations the smallest distance between two points cannot be
  smaller than the grid spacing. Thus the correlations are binned,
  each bin covering a range $\Delta r = 0.075d$. A characteristic
  result is shown in Figure \ref{fig:lcorr}, where the correlations
  from runs with different Coriolis numbers at the southern pole are
  presented. The main conclusion is that the correlation diminishes
  faster as function of $r$ when rotation is more rapid, implying that
  $l_{\rm c}$ decreases as a function of rotation. There is, however,
  a problem of how exactly to define $l_{\rm c}$. Due to the binning
  the average correlation in the bin with the smallest $r$ varies from
  one calculation to the other, and setting a fixed threshold value
  would be more arbitrary than in the case of the correlation time. We
  see that a reasonable solution to bypass this problem is to use test
  particles in order to determine the turnover time, $t_{\rm to}
  (\propto l_{\rm c}/u_{\rm t})$, that captures the changing spatial
  scale of convection as a function of rotation.

\subsection{Turnover time}
\label{subsec:turnovert}
  The turnover time is usually estimated by dividing a characteristic
  length scale by a characteristic velocity. For convection
  calculations one often uses the depth of the convection zone divided
  by the volume-averaged rms velocity, i.e.
  \begin{equation}
    t_{\rm to}^{(\rm s)} = \frac{d}{u_{\rm t}}\;,
    \label{equ:tausimple}
  \end{equation}
  where the superscript s refers to `simple estimate'. For example,
  for the Co0 run, $u_{\rm t} \approx 0.085$ and $d = 1$, giving
  $t_{\rm to}^{(\rm s)} \approx 12.0$. Using this estimate with
  $\tau_{\rm c} = 9.4$ would indicate that St is approximately
  0.8. Although this simple method probably gives the correct order of
  magnitude of the turnover time, it is still quite a crude estimate
  since the values for the typical scales and velocities are rather
  uncertain and vary nonlinearly as function of rotation and Rayleigh
  number (see below). Thus a more precise way of calculation is
  desirable.

  \begin{figure}
  \centering
  \includegraphics[width=0.5\textwidth]{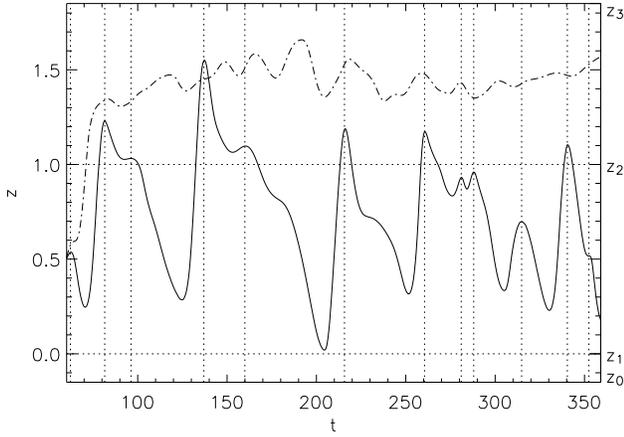}
     \caption{Vertical positions of two test particles in run the
       Co0. The dotted vertical lines denote the times where the
       particle denoted by the solid line changes direction from
       downward motion to upward motion. The dash-dotted line shows a
       particle which is stuck in the lower overshoot layer for the
       whole duration of the calculation.}
     \label{fig:etestp}
  \end{figure}

  \begin{figure}
  \centering
  \includegraphics[width=0.5\textwidth]{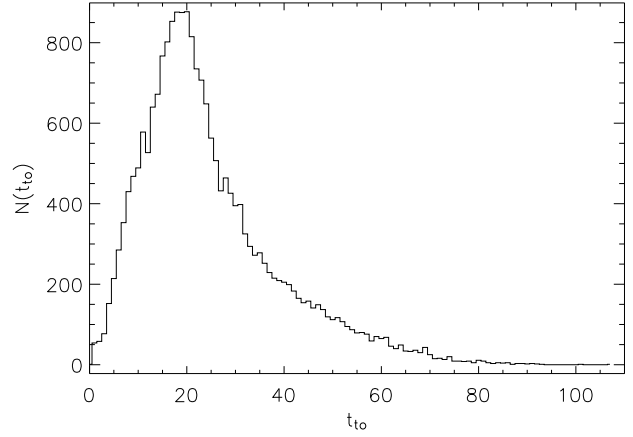}
     \caption{Distribution of turnover times from run the Co0 with the
       test particle method.}
     \label{fig:disto}
  \end{figure}

  \begin{table}
    \centering
    \caption[]{From left to right: rms-velocity averaged over the
      convectively unstable region and time, correlation time from the
      autocorrelation of the vertical velocity, turnover time from
      test particle trajectories, and the Strouhal number. The last
      column states the number of snapshots with respect to which the
      correlation times and lengths were calculated.}
    \vspace{-0.25cm}
    \label{tab:strouhal}
     $$
    \begin{array}{p{0.15\linewidth}ccccccc}
      \hline
      \noalign{\smallskip}
      Run       & u_{\rm t} & \tau_{\rm c} & t_{\rm to} & {\rm St} & N \\
      \noalign{\smallskip}
      \hline
      lCo0      & 0.085 & 11.2 \pm 2.3 & 25.9 & 0.43 & 209 \\
      Co0       & 0.084 &  9.4 \pm 2.1 & 25.1 & 0.37 & 206 \\
      mCo0      & 0.080 &  8.8 \pm 2.6 & 24.0 & 0.37 & 212 \\
      hCo0      & 0.073 &  8.9 \pm 1.6 & 23.4 & 0.38 & 199 \\
      \hline
      Co01-00   & 0.084 &  9.4 \pm 2.2 & 25.0 & 0.38 & 206 \\
      Co01-30   & 0.084 &  9.2 \pm 1.5 & 25.1 & 0.37 & 202 \\
      Co01-60   & 0.082 &  9.8 \pm 2.9 & 25.0 & 0.39 & 207 \\
      Co01-90   & 0.082 &  9.0 \pm 1.6 & 25.5 & 0.35 & 203 \\
      Co1-00    & 0.083 &  7.3 \pm 1.4 & 24.5 & 0.30 & 207 \\
      Co1-30    & 0.086 &  6.1 \pm 0.9 & 23.4 & 0.26 & 198 \\
      Co1-60    & 0.084 &  6.9 \pm 0.8 & 23.7 & 0.29 & 207 \\
      Co1-90    & 0.083 &  7.1 \pm 0.9 & 23.8 & 0.30 & 206 \\
      Co10-00   & 0.201 &  2.2 \pm 0.3 & 12.6 & 0.17 & 202 \\
      Co10-30   & 0.072 &  3.9 \pm 0.3 & 21.0 & 0.19 & 200 \\
      Co10-60   & 0.062 &  3.4 \pm 0.2 & 20.5 & 0.17 & 201 \\
      Co10-90   & 0.062 &  3.1 \pm 0.4 & 18.4 & 0.17 & 201 \\
      \noalign{\smallskip}
      \hline
    \end{array}
    $$ 
  \end{table}

  A way to improve the estimate given by Eq.~(\ref{equ:tausimple}) is
  to follow the trajectories of Lagrangian test particles in the flow.
  This is done by finding the times where a particle changes its
  direction, i.e. turns over. Thus, one turnover time would be the
  time between two consecutive turns to the same direction, e.g. from
  downward motion to upward motion. The advantage of this method is
  that the assumption of the vertical scale of convection is removed.
  However, the danger with this method is that the smallest scales
  begin to dominate due to, for example, contributions from particles
  stuck in the stable layer. This problem is discussed below.

  Figure \ref{fig:etestp} shows the trajectories of two test particles
  in the run Co0. The solid line shows a particle which is carried by
  the convective flow throughout the calculation (the particles were
  introduced in the flow at $t = 60$). In the figure we only denote
  the turnovers from downward to upward motion, but including also the
  opposite changes of direction a total of 21 turnovers are
  registered. Figure \ref{fig:disto} shows a histogram of the
  registered turnover times from run Co0. In total, the thousand
  particles make 21300 individual turnovers in the course of the
  calculation. The distribution is centered near 20 having an
  exponential tail towards longer times. Arithmetic average over the
  distributions shown in Figure \ref{fig:disto} gives a turnover time
  $t_{\rm to} \approx 25$, which indicates that the Strouhal number to
  be significantly less than 0.8 which was obtained with the simple
  estimate of the turnover time.
  
  The test particle method takes into account the variable spatial
  scale of convection but it also picks up the small scale turnovers
  in the stable layer. A small number of particles get stuck in the
  lower overshoot layer for a long time and some even stay there for
  the complete duration of the calculation (the dash-dotted line in
  Figure \ref{fig:etestp}) and contribute a large number of small
  turnover times. These particles, however, have only a small effect
  on the estimate of $t_{\rm to}$ for the nonrotating and slowly
  rotating cases where the turnover time might be slightly
  underestimated.

  From Figure \ref{fig:lcorr} and the second column of Table
  \ref{tab:strouhal} we can determine that whilst $l_{\rm c}$
  decreases as a function of rotation, a similar trend is visible in
  $u_{\rm t}$. Thus we would expect that the ratio $u_{\rm t}/l_{\rm
    c}$ not to vary much as a function of Co. This is indeed the trend
  that we observe for $t_{\rm to}$ which seems to support the
  conjecture that $t_{\rm to } \propto l_{\rm c}/u_{\rm t}$.

  \begin{figure*}
  \centering
  \includegraphics[width=1.\textwidth]{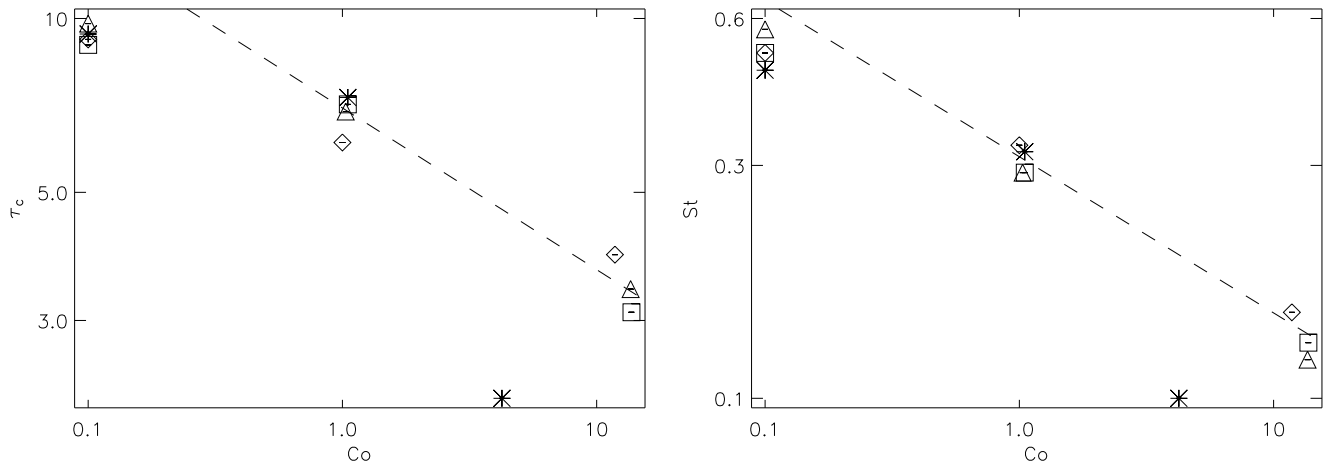}
     \vspace{-6.5cm}
     \caption{The correlation time $\tau_{\rm c}$ (left) and the
       Strouhal number ${\rm St}$ (right) functions of the Coriolis
       number. Powerlaws ${\tau_{\rm c} \propto {\rm Co}^{-0.28}}$ and
       ${\rm St} \propto {\rm Co}^{-0.32}$ are plotted. The stars,
     diamonds, triangles, and squares represents the calculations at
     latitudes $0 \degr$ (equator), $-30\degr$, $-60\degr$, and $-90
     \degr$ (south pole), respectively.}
     \label{fig:st}
  \end{figure*}

  \subsubsection{Dependence on the Rayleigh number}
  We have made a set of runs varying the Rayleigh number from $1.25
  \cdot 10^5$ to $10^6$, denoted by lCo0, Co0, mCo0, and hCo0 in Table
  \ref{tab:strouhal}. With the test particle method $t_{\rm to}$
  decreases slightly as a function of Ra which can be explained by the
  fact that contributions of smaller scales are now resolved in the
  model leading to a shorter correlation length, indications of which
  are also visible in the velocity autocorrelations. However, we do
  not find any significant trend in the correlation time as function
  of Ra. It follows that with the present definition the Strouhal
  number saturates a value approximately 0.4. This behaviour is
  similar to the Reynolds number independence of the Strouhal number
  found by Brandenburg et al. (\cite{Brandea04}).

  \subsubsection{Effects of rotation}
  \label{subsec:strot}
  In Paper I, we have shown that the efficiency of convection and
  overshooting is reduced as rotation increases. The latter effect
  indirectly indicates that the spatial scale of convection decreases,
  which could lead to shorter turnover time if the overall velocities
  are less affected. The decreasing horizontal scale of convection as
  a function of rotation can be seen for example in Figure 4 of Paper
  I. In order to study the effects of rotation on the Strouhal number
  we have made three sets of calculations with varying rotational
  influence. We denote these runs by the prefix Co01, Co1, and Co10,
  which correspond to the approximate Coriolis numbers 0.1, 1, and 10
  in the runs (the actual values of Co achieved in the calculations
  vary, see Table \ref{tab:convruns}). We also probe the latitudinal
  dependence by making calculations at latitudes $0\degr$, $-30\degr$,
  $-60\degr$, and $-90\degr$ for each rotation rate. The correlation
  and turnover times, and the resulting Strouhal numbers for all the
  calculations are presented in Table \ref{tab:strouhal}.

  For the slowest rotation, we find little differences to the
  nonrotating case Co0 which was discussed above. Also the variation
  of the timescales as function of latitude is small in comparison to
  the variation within individual runs. For the Co1 set the
  correlation time is markedly shorter (about 6-7 time units as
  opposed to 9-10 in the run Co0 and the Co01 set). This can be
  explained by the deflection of the vertical flows by the Coriolis
  force, leading to smaller vertical scale of convection which tends
  to shorten the correlation time. Support for this conjecture is
  given by the decrease of the turnover time from the test
  particles. A similar decrease in correlation time is noted to occur
  also if it is estimated from the horizontal velocities.

  For the most rapidly rotating case, Co10, the trend of decreasing
  spatial scales continues. This is manifested in the clearly shorter
  turnover time $t_{\rm to}$ as opposed to the more slowly rotating
  runs discussed above. Here, it is also important to note the
  misleading value of the turnover time given by the simple method,
  Eq.~(\ref{equ:tausimple}). Whereas $t_{\rm to}^{(s)}$ increases with
  rotation due to the smaller velocities in general, the actual
  turnover time decreases due to the smaller spatial scale of
  convection. As for the correlation time, the strong Coriolis forces
  tend to disrupt the cellular structure of the convection rapidly
  resulting in a shorter $\tau_{\rm c}$. Considering the Strouhal
  number, the decrease in $\tau_{\rm c}$ overweights the decrease of
  the turnover time so that the actual value of St decreases as well.

  We find that St is more or less consistent with a powerlaw ${\rm St}
  \propto {\rm Co}^{-0.32}$ as a function of the Coriolis number for
  moderate and rapid rotation (see Figure \ref{fig:st}).

  \subsection{Implications for first order smoothing}
  The relatively large values of St raise question of the validity of
  the first order smoothing approximation. To our knowledge, this
  question has been considered only by NS88, who derive a fourth order
  correlation approximation from which they derive the transport
  coefficient responsible for the generation ($\alpha$-effect) and
  diffusion of the mean magnetic field for homogenuous and isotropic
  turbulence for which the velocity field and the correlation time is
  known. Their results indicate that the fourth order contributions,
  become comparable to the second order effects, i.e. the FOSA result,
  if the Strouhal number is of the order 0.5 or larger and that the
  cumulative expansion fails to converge if St $>$ 1. In this light
  our results would indicate that one should be cautious in applying
  FOSA in the interpretation of the present convection
  calculations. Direct testing of this, however, requires that we
  compare analytical mean-field expressions of the relevant transport
  coefficients to the numerical results. Preliminary results indicate
  that the correlation time needed to fit the numerical data is
  significantly shorter than that measured from the velocity
  autocorrelation function (see K\"apyl\"a et al. \cite{Kapyea05}).

\section{Conclusions}
\label{sec:conclu}
  We estimate the nondimensional measure of the correlation time,
  the Strouhal number, from numerical models of convection. We
  calculate the correlation and turnover times separately from the
  velocity autocorrelation function and the trajectories of embedded
  test particles, respectively, and define the Strouhal number as the
  ratio of the two.

  The Strouhal number arises in the mean-field theories of angular
  momentum transport and hydromagnetic dynamos where its value
  determines the validity of certain widely used approximations, such
  as the first order smoothing. These approximations are based on a
  cumulative series expansion of the relevant turbulent correlation,
  e.g. the electromotive force in the dynamo theory. Essentially, the
  higher order terms in this expansion are proportional to the
  Strouhal number. Thus the value of St determines whether or not the
  expansion converges.

  The main results can be summarised as follows:
  \begin{itemize}
    \item We find that the correlation time does not depend on the
      Rayleigh number for the parameter range explored in the present
      study. Note that the results for $\tau_{\rm c}$ presented here
      differ significantly from those of K\"apyl\"a et
      al. (\cite{Kapyea05}) due to an error in the analysis in the
      latter study.

    \item As function of rotation, $\tau_{\rm c}$ decreases by a
      factor of roughly three when the Coriolis number is increased
      from 0 to 10. At the same time, the turnover time decreases only
      by about $20\%$, leading to an approximate relation ${\rm
        St} \propto {\rm Co}^{-0.32}$ for moderate and rapid rotation.

    \item A noteworthy fact is that if one takes the usual estimate of
      dividing the characteristic length by the characteristic
      velocity, the turnover time increases as function of rotation
      due to the overall decreasing velocities. However, the test
      particle data gives an opposite result. The discrepancy is due
      to the fact that even though velocities decrease in general, the
      spatial scale of convection decreases even more. Thus it is
      imperative that the definition of St takes into account the
      spatial scale of the motions, a fact which was also noted by
      NS88.
  \end{itemize}
   
  Our inability to find any dependence on the Rayleigh number and the
  relatively high values ($0.1 \ldots 0.4$) of the Strouhal number
  raise the question of the validity of FOSA. In the study of NS88,
  although not from a directly comparable context, the higher order
  effects became visible when St was of similar magnitude. At the
  moment, we cannot draw a firm conclusion whether or not it is safe
  to use FOSA or not. We think that the next step in the direction of
  determining this would be to develop mean-field expressions for the
  transport coefficients that are directly applicable to the present
  numerical calculations. Furthermore, these expressions should be
  derived using FOSA and preferably also with some higher order
  expansion in order to determine which of them gives a better
  description of the numerical result, and, more importantly, whether
  the whole concept of the cumulative expansion is applicable. This
  study, however, does not fit in the scope of the present paper.

\begin{acknowledgements}
  PJK acknowledges the financial support from the Finnish graduate
  school of astronomy and space physics and travel support from the
  Kiepenheuer intitute. PJK thanks NORDITA and its staff for their
  hospitability during his visit. MJK acknowledges the hospitality of
  LAOMP, Toulouse and the Kiepenheuer-Institut, Freiburg during her
  visits, and the Academy of Finland project 203366. Travel support
  from the Academy of Finland grant 43039 is acknowledged. The authors
  thank Axel Brandenburg and Michael Stix for their useful comments on
  the manuscript and Wolfgang Dobler for illuminating discussions. The
  anonymous referee is acknowledged for the helpful comments on the
  manuscript.
\end{acknowledgements}

\end{document}